\documentclass[10pt,letterpaper,twocolumn]{article} 

\usepackage{ol2}
\usepackage[draft]{hyperref}
\usepackage{amsmath}

\begin{document}

\twocolumn[ 

\title{Tomographic imaging of asymmetric molecular orbitals with a two-color multicycle laser field }

\author{Meiyan Qin$^1$, Xiaosong Zhu$^{1}$, Qingbin Zhang$^{1,2,*}$, and Peixiang Lu$^{1,2}$$^*$}

\address{$^1$Wuhan National Laboratory for Optoelectronics and School of Physics, Huazhong University of Science and Technology, Wuhan 430074, China \\
$^2$MOE Key Laboratory of Fundamental Quantities
Measurement,Wuhan430074,China\\
} \email{$^*$Corresponding authors: zhangqingbin@mail.hust.edu.cn,
lupeixiang@mail.hust.edu.cn}

\begin{abstract}
We theoretically demonstrate a scheme for tomographic
reconstruction of asymmetric molecular orbitals based on
high-order harmonic generation with a two-color multicycle laser
field. It is shown that by adjusting the relative phase of the two
fields, the returning electrons can be forced to recollide from
one direction for all the orientations of molecules. Thus the
reconstruction of the asymmetric orbitals can be carried out with
multicycle laser field. This releases the stringent requirement of
a single-cycle pulse with a stabilized and controllable
carrier-envelop phase for the tomographic imaging of asymmetric
molecular orbitals.
\end{abstract}

] 

\noindent The recent developments in strong-field physics make it
possible to probe the molecular structure and electron dynamics
with attosecond and $\mathrm{\AA}$ngstr\"{o}m resolutions
\cite{haessler1,sali,worner,zhou}. As a potential tool of imaging
the molecular electron dynamics, molecular orbital tomography
(MOT) based on high-order harmonic generation (HHG) has attracted
a great deal of attention in the past
decade\cite{itatani,haessler2,levesque,vozzi,zwan1}. This
tomographic method was first proposed and successfully carried out
by J.Itatani {\it et al.} \cite{itatani} to reconstruct the
highest occupied molecular orbital (HOMO) of N$_2$ in experiment.
Since then, much effort has been expended to extend the MOT method
to molecules other than N$_2$ \cite{vozzi,zwan1}. It has been
shown that the MOT method can be directly applied to the symmetric
molecular orbitals. While for asymmetric molecular orbitals, to
reconstruct the orbitals from HHG, the returning wave-packets
should be controlled to recollide with the parent ion from only
one direction \cite{zwan1}.

It was proposed that the unidirectional recollisions can be
achieved by using an extremely short tailored laser pulse
\cite{zwan1}. In this scheme, by stabilization and control of the
carrier-envelope phase (CEP), the returning wave-packet can be
controlled. However, this scheme requires a single-cycle pulse
with a stabilized and controllable carrier-envelop phase, which is
a rather stringent requirement. This kind of laser pulse with a
sufficient intensity is still not available for many laboratories.
Therefore, a method with less stringent experimental conditions to
achieve the unidirectional recollisions is preferred for
tomographic reconstruction of asymmetric molecular orbitals. It
has been shown that the two-color laser field is an efficient tool
to control the electron dynamics. For example, a tailored
two-color laser field has been used to control the electron
dynamics for producing attosecond pulses \cite{lan,cao}. In this
Letter, we theoretically demonstrate a method for tomographic
imaging of asymmetric molecular orbitals with a multicycle
two-color laser pulse. By adjusting the relative phase of the two
fields, the returning wave-packets are forced to recollide from
only one direction for all the orientations of the molecule.
Following the tomographic procedure, the asymmetric orbital is
satisfactorily reconstructed.

In our scheme, we use the multicycle two-color laser field to
control the return of the electron. To characterize the returning
wave-packet, the probability that an electron returns with
momentum $k$ is calculated semi-classically \cite{zwan2}. In
detail, the tunnel ionization rate at the ionization time $t_i$,
is calculated within the Molecular Ammosov-Delone-Krainov (MO-ADK)
model for oriented molecules \cite{tong}. After tunnelling, the
electron follows a classical trajectory. Every classical return
receives a weight based on the tunnelling probability and a factor
$\tau^{-3}$, where $\tau$ is the time the electron spends in the
continuum until the time of return and reflects the effect of
wave-packet spreading. We take the example of CO to demonstrate
the efficient control of the recollision wave-packet by the
two-color laser field. A 20-fs (FWHM) 1600-nm linearly polarized
pulse with an intensity of $1.0\times10^{14}W/cm^2$ and a 20-fs
800-nm linearly polarized pulse with an intensity of
$9.0\times10^{12}W/cm^2$ are used to synthesize the two-color
driving field. The electric field is given by
$E(t)=E_0f(t)cos(w_0t)+E_1f(t)cos(2w_0t+\phi)$, where $E_0$, $E_1$
and $\omega_0$, $2\omega_0$ are the amplitudes and the frequencies
of the 1600-nm and 800-nm pulses, respectively. $\phi$ is the
relative phase between the two pulses, which can be easily
adjusted in experiment. The envelope $f(t)$ is a sine squared
function.
\begin{figure}[htb]
\centerline{\includegraphics[width=8.0cm]{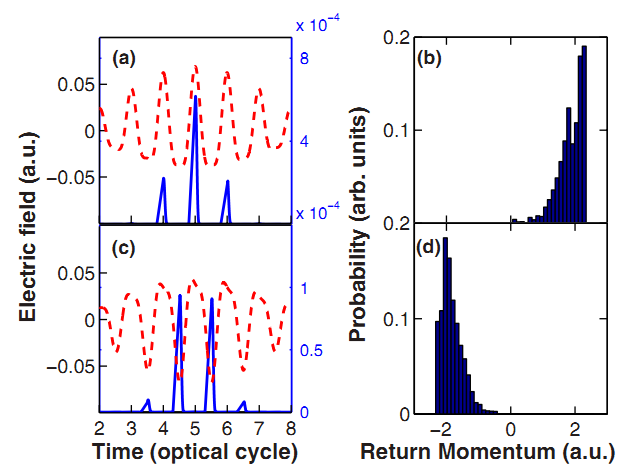}}
\setlength{\abovecaptionskip}{-0.1cm}
\setlength{\belowcaptionskip}{-0.15cm} \caption{Ionization rate as
a function of the ionization time (blue solid line) for the
two-color field with $\phi=0.0\pi$ (panel a) and $\phi=0.9\pi$
(panel c). The corresponding electric fields are also depicted by
the red dashed line. The probability that an electron returns with
momentum $k$ for $\phi=0.0\pi$ and $\phi=0.9\pi$ are presented in
panel b and panel d, respectively.}
\end{figure}

In figure 1, both the ionization rate (left column) and the
probability that an electron returns with momentum $k$ (right
column) are depicted for two different relative phases
$\phi=0.0\pi$ (the first row) and $\phi=0.9\pi$ (the second row).
For each pulse the sum of the probabilities is normalized to 1.
The CO molecule is oriented at $0^\circ$. The corresponding
electric fields are also displayed by the red dashed lines in this
figure. As shown in the first row for $\phi=0.0\pi$, the tunnel
ionization of the electron is confined around the peaks of the
electric field with positive amplitude. Correspondingly, the
electrons recollide with positive momentum, as shown in panel b.
In second row for $\phi=0.9\pi$, the electron is ionized around
the peaks with negative amplitude and then all the electrons
return with negative momentum. For both cases, the returning
electrons are controlled to approach the parent ion from one side,
which is also suggested in ref. \cite{haessler1}.

\begin{figure}[htb]
\centering\includegraphics[width=8cm]{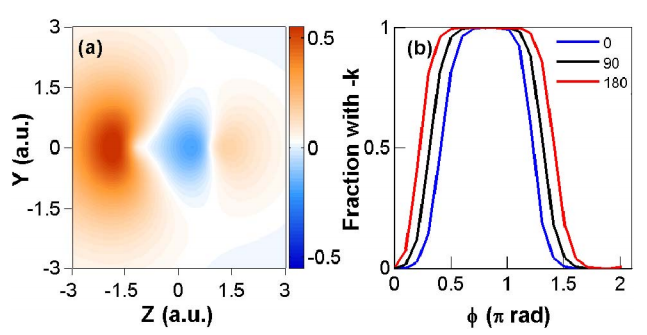}
\setlength{\abovecaptionskip}{-0.1cm}
\setlength{\belowcaptionskip}{-0.05cm}\caption{(a) A normalized
two-dimensional projection of the highest occupied molecular
orbital of CO obtained from the Gaussian 03 $ab$ $initio$ code.
The orientation angle is at $0^\circ$. (b) Fraction of the
electron returning with negative momentum as a function of the
relative phase $\phi$ for three different orientations of CO
molecule. The blue, black and red lines correspond to the
orientation angle $0^\circ$, $90^\circ$ and $180^\circ$,
respectively. }
\end{figure}

To further demonstrate the efficient control of the returning
electron by the two-color fields, the fraction of the electrons
with negative momentum as a function of the relative phase $\phi$
is presented in Fig. 2. The wavefunction of CO from the Gaussian
03 $ab$ $initio$ code \cite{gauss} is also depicted in Fig. 2a,
where the orientation angle is $0^\circ$. According to the MOT
theory, the unidirectional recollision is required for all the
orientations of the molecule. Therefore, the momentum distribution
of the returning electron for other orientations are also
investigated. The results for molecule CO oriented at $0^\circ$
(blue line) , $90^\circ$ (black line) and $180^\circ$ (red line)
are presented in Fig. 2(b). As shown in Fig. 2(b), by adjusting
the relative phase of the two fields, the two-color multicycle
laser pulse can provide an efficient control of the returning
wave-packet. For the cases $0.7\pi\leq\phi\leq1.0\pi$, the
electron returns are well confined to only one direction for all
the orientations. This offers the possibility of the
reconstruction of asymmetric molecular orbitals with multicycle
laser pulse.

\begin{figure}[htb]
\centering\includegraphics[width=8cm]{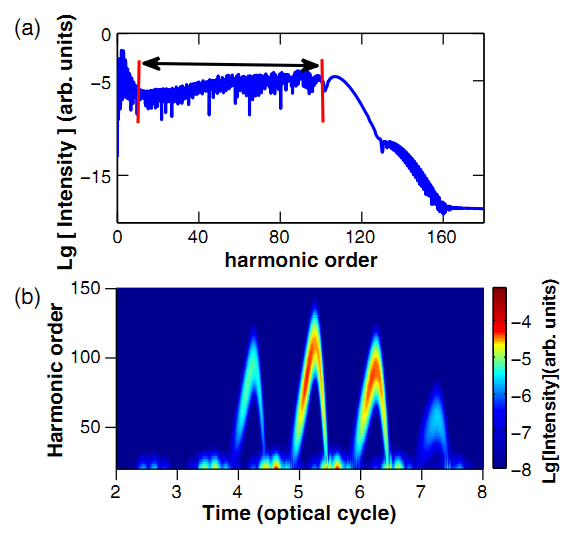}
\setlength{\abovecaptionskip}{0.001cm}
\setlength{\belowcaptionskip}{-0.12cm} \caption{(a) The harmonic
spectrum of CO oriented at the angle $0^\circ$ for the two-color
field with $\phi=0.9\pi$. The double arrow indicates the spectral
range (harmonics 11 to 101) sampled for orbital reconstruction.
(b) The corresponding time-frequency distribution of
HHG.}\vspace*{-0.1cm}
\end{figure}

In the following, we choose the two-color multicycle laser fields
with $\phi=0.9\pi$ to carry out the reconstruction of the HOMO of
CO. The reference Kr atom with the same ionization potential
(0.519 a.u.) as that of the HOMO of CO is used \cite{tong}. The
complex amplitudes of the high-order harmonics are calculated with
the SFA model \cite{lewen}. In Fig. 3, the high-order harmonic
spectrum (panel a) and the corresponding time-frequency
distribution of HHG (panel b) are presented for the orientation
angle $0^\circ$. From Fig. 3(b), one can see that the emission of
the high-order harmonics are well confined to the time when the
direction of the electric field is positive. Therefore, the
wave-packets approach the core mainly from one side, which is
consistent with the result obtained within semi-classical model as
shown in the second row of Fig. 1.

\begin{figure}[htb]
\centering\includegraphics[width=8cm]{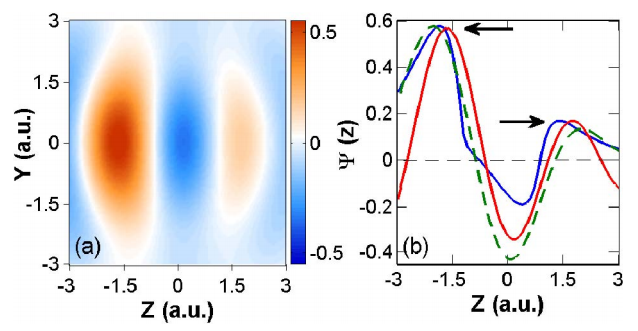}
\setlength{\abovecaptionskip}{-0.1cm}
\setlength{\belowcaptionskip}{-0.2cm} \caption{(a) The normalized
real part of the reconstructed orbital of the HOMO of CO using
multicycle two-color laser field with $\phi=0.9\pi$. (b) Cuts
along the internuclear axis for the $ab$ $initio$ orbital (blue
line), the reconstructed orbital (red line) and the
correspondingly Fourier-filtered $ab$ $inito$ orbital (dashed
green line).}
\end{figure}

To reconstruct the HOMO of CO, the odd harmonics 11 to 101 at 19
different angles between 0 and 180$^\circ$ are sampled. Following
the reconstruction procedure in the velocity form based on the
plane wave approximation\cite{haessler1}, the HOMO of CO is
reconstructed by using the dipole matrix elements projected
perpendicular to the internuclear axis. In Fig. 4(a), the
normalized real part of the reconstructed orbital is presented
with the same color scale as that of Fig. 2(a). As shown in Fig.
2(a) and Fig. 4(a), the structure of the HOMO of CO is well
reproduced with the multicycle two-color laser field. In detail,
the reconstructed orbital possesses three main lobes with
alternating signs, separated by two nodal surfaces. For clarity,
cuts along the internuclear axis for the reconstructed orbital
(red line) and the $ab$ $initio$ orbital (blue line) are depicted
in Fig. 4(b). As shown in this figure, the two positive maxima of
the reconstructed and $ab$ $initio$ orbital have the same values,
as indicated by the two black arrows. Whereas some distortions are
also observed in the reconstructed orbital. For example, the
detailed structure near the left nucleus is lost and the amplitude
at the negative maximum is lower than that of the exact
wavefunction. To find the origin of these distortions, we
calculate the Fourier-filtered $ab$ $inito$ orbital with the
$\mathbf{k}$-range corresponding to the sampled spectral range. A
cut along the internuclear axis for the Fourier-filtered $ab$
$inito$ orbital is depicted by the dashed green line in Fig. 4b.
One can see that the same deviations are observed for the
Fourier-filtered $ab$ $inito$ orbital. Therefore, these
distortions mainly originate from the limited spectral range, of
which the effect on the reconstruction has been discussed in
detail in \cite{haessler1,vozzi}. Besides, there is a slight shift
of the reconstructed wavefunction relative to the exact
wavefunction and the origin of the shift is still an open
question. Although only the HOMO of CO is considered in our
simulation, our scheme can be extended to other asymmetric
molecular orbitals by properly choosing the relative phase and
intensity of the two-color fields. Experimentally, there still
exist some challenges such as the orientation of the molecules
\cite{frumker} and the measurements of the harmonic phase and
polarization, which is systematically discussed in
\cite{haessler1,sali}.

In summary, we theoretically demonstrate a method for tomographic
reconstruction of asymmetric molecular orbitals with multicycle
two-color laser pulses. The unidirectional recollision of the
electron wave-packet is achieved for all the orientations of the
molecules by adjusting the relative phase of the two fields, and
then the asymmetric molecular orbtal is satisfactorily
reconstructed with multicycle laser pulses. This releases the
stringent requirement of a single-cycle pulse with a stabilized
and controllable carrier-envelop phase for the tomographic imaging
of asymmetric molecular orbitals.

This work was supported by the NNSF of China under Grants No.
11234004 and 60925021, the 973 Program of China under Grant No.
2011CB808103 and the Doctoral fund of Ministry of Education of
China under Grant No. 20100142110047.

\newpage
\twocolumn


\begin{thebibliography}{99}

\bibitem{haessler1} S. Haessler, J. Caillat and P. Sali\`{e}res, ``Self-probing of molecules with high harmonic generation,'' J.
Phys. B: At. Mol. Opt. Phys. {\bf 44}, 203001 (2011).

\bibitem{sali} P. Sali\`{e}res, A. Maquet, S. Haessler, J. Caillat
and R. Ta\"{\i}eb, ``Imaging orbitals with attosecond and
$\mathrm{\AA}$ngstr\"{o}m resolutions: toward attochemistry?,''
Rep. Prog. Phys. {\bf 75},062401 (2012).

\bibitem{worner} H. J. W\"{o}rner, J. B. Bertrand, D. V.
Kartashov, P. B. Corkum and D. M. Villeneuve, ``Following a
chemical reaction using high-harmonic interferometry,'' Nature
{\bf 466}, 604 (2010).

\bibitem{zhou} Y. Zhou, C. Huang, Q. Liao and P. Lu, ``Classical simulations including electron correlations for sequential double ionization,'' Phys. Rev. Lett. {\bf
109}, 053004 (2012).

\bibitem{itatani} J. Itatani, J. Levesque, D. Zeidler, H. Niikura,
H. P\'{e}pin, J. C. Kieffer, P. B. Corkum and D. M. Villeneuve,
``Tomographic imaging of molecular orbitals,'' Nature {\bf 432},
867 (2004).

\bibitem{haessler2} S. Haessler, J. Caillat, W. Boutu,
C. Giovanetti-Teixeira, T. Ruchon, T. Auguste, Z. Diveki, P.
Breger, A. Maquet, B. Carr\'{e}, R. Ta\"{\i}eb and P.
Sali\`{e}res, ``Attosecond imaging of molecular electronic
wavepackets,'' Nature Phys, {\bf 6}, 200 (2010).

\bibitem{levesque} J. Levesque, D. Zeidler, J. P. Marangos, P. B.
Corkum and D. M. Villeneuve, ``High harmonic generation and the
role of atomic orbital wave functions,'' Phys. Rev. Lett. {\bf
98}, 183903 (2007).

\bibitem{vozzi} C. Vozzi, M. Negro, F. Calegari, G. Sansone, M.
Nisoli, S. De Silvestri and S. Stagira, ``Generalized molecular
orbital tomography,'' Nature Phys, {\bf 7}, 822 (2011).

\bibitem{zwan1} E. V. Zwan, C. C. Chiril$\check{a}$ and M. Lein, ``Molecular orbital tomography using short laser pulses,'' Phys. Rev. A {\bf 78}, 033410 (2008).

\bibitem{lan} P. Lan, P. Lu, W. Cao, Y. Li, and X. Wang, ``Isolated sub-100-as pulse generation via controlling electron dynamics,'' Phys. Rev. A {\bf
76}, 011402 (2007).

\bibitem{cao} W. Cao, P. Lu, P. Lan, X. Wang and G.
Yang, ``Efficient isolated attosecond pulse generation from a
multi-cycle two-color laser field,'' Opt. Express {\bf 15}, 530
(2007).

\bibitem{zwan2} E. V. Zwan and M. Lein, ``Control of recollision wave packets for molecular orbital tomography using short laser pulses,'' J.
Phys. B: At. Mol. Opt. Phys. {\bf 41}, 074009 (2008).

\bibitem{tong} X. M. Tong, Z. Zhao and C. D. Lin, ``Theory of molecular tunneling ionization,'' Phys. Rev. A {\bf 66},
033402 (2002).

\bibitem{gauss} M. J. Frisch, G. W. Trucks, H. B. Schlegel, G. E. Scuseria, M. A. Robb, J. R. Cheeseman, J. A. Montgomery Jr.,
T. Vreven, K. N. Kudin, J. C. Burant, J. M. Millam, S. S. Iyengar,
J. Tomasi, V. Barone, B. Mennucci, M. Cossi, G. Scalmani, N. Rega,
G. A. Petersson, H. Nakatsuji,M. Hada,M. Ehara, K. Toyota, R.
Fukuda, J. Hasegawa,M. Ishida, T. Nakajima, Y. Honda, O. Kitao, H.
Nakai, M. Klene, X. Li, J. E. Knox, H. P. Hratchian, J. B. Cross,
V. Bakken, C. Adamo, J. Jaramillo, R. Gomperts, R. E. Stratmann,
O. Yazyev, A. J. Austin, R. Cammi, C. Pomelli, J. W. Ochterski, P.
Y. Ayala, K. Morokuma, G. A. Voth, P. Salvador, J. J. Dannenberg,
V. G. Zakrzewski, S. Dapprich, A. D. Daniels, M. C. Strain, O.
Farkas, D. K. Malick, A. D. Rabuck, K. Raghavachari, J. B.
Foresman, J. V. Ortiz, Q. Cui, A. G. Baboul, S. Clifford, J.
Cioslowski, B. B. Stefanov, G. Liu, A. Liashenko, P. Piskorz, I.
Komaromi, R. L. Martin, D. J. Fox, T. Keith, M. A. Al-Laham, C. Y.
Peng, A. Nanayakkara, M. Challacombe, P. M. W. Gill, B. Johnson,
W. Chen, M. W. Wong, C. Gonzalez, and J. A. Pople, ¡°Gaussian 03,
Revision C.02,¡± Gaussian Inc., Wallingford, CT (2010).

\bibitem{lewen} M. Lewenstein, Ph. Balcou, M. Yu. Ivanov, A. L'Huillier,
and P. Corkum, ``Theory of high-harmonic generation by
low-frequency laser fields,'' Phys. Rev. A {\bf 49,} 2117 (1994).

\bibitem{frumker} E. Frumker, C. T. Hebeisen, N. Kajumba, J. B.
Bertrand, H. J. W$\ddot{o}$rner, M. Spanner, D. M. Villeneuve, A.
Naumov and P. B. Corkum, ``Oriented rotational wave-packet
dynamics studies via high harmonic generation,'' Phys. Rev. Lett.
{\bf 109}, 113901 (2012)


\end{thebibliography}
\end{document}